# The generalized Kochen-Specker theorem

P.K.Aravind
Physics Department, Worcester Polytechnic Institute, Worcester, MA 01609

## ABSTRACT

A proof of the generalized Kochen-Specker theorem in two dimensions due to Cabello and Nakamura is extended to all higher dimensions. A set of 18 states in four dimensions is used to give closely related proofs of the generalized Kochen-Specker, Kochen-Specker and Bell theorems that shed some light on the relationship between these three theorems.

Cabello and Nakamura [1] recently proved a generalization of the Kochen-Specker (KS) theorem [2] in two dimensions suggested by a similar generalization of Gleason's theorem [3]. The key element in the generalization of both theorems is the replacement of von Neumann (or projective) measurements by more general measurements described by positive operator valued measures (POVMs). This shift leads to two unexpected, but pleasing, effects: it causes the validity of both theorems to extend down to dimension two (rather than three as previously), and it also makes the proofs of both theorems much easier. The purpose of this note is, firstly, to show how the proofs of the generalized Kochen-Specker (GKS) theorem due to Cabello and Nakamura can be extended to arbitrary finite dimension $d \geq 2$, and, secondly, to exhibit a set of closely related proofs of the GKS, KS and Bell theorems that sheds some light on the mutual relationship of these three theorems.

A POVM consists of a set of positive semi-definite operators providing a resolution of the identity. The generalized measurement corresponding to a POVM can be realized by coupling the system to an ancillary system and carrying out a von Neumann measurement on the enlarged system; this measurement then appears in the space of the system alone as a POVM. The GKS theorem states that in a Hilbert space of dimension two or greater it is always possible to find a finite set of positive semi-definite operators ("elements" [4]) that cannot each be assigned the value 0 or 1 in such a way that any subset of elements constituting a POVM contains exactly one element with the value 1.

To prepare the way for the generalization in the next paragraph, we first review Nakamura's proof of the GKS theorem in dimension 2 [1]. Consider the spin-up states of a spin-½ particle along the six directions from the center of a regular hexagon to its vertices. Choose as POVM elements the six projectors onto these states, each multiplied by a factor of ½. The four elements associated with any two pairs of opposite directions then constitute a POVM, there being three such POVMs altogether. Suppose now that it is possible to assign a 0 or 1 to every element in such a way that each POVM has exactly one element with the value 1 in it. Then we can deduce the following contradiction: on the one hand, the total number of occurrences of elements with value 1 in all three POVMs is odd (because there is exactly one such element per POVM) but, on the other hand, it is also even (because each element with value 1 occurs in two POVMs and hence is counted twice). This contradiction shows the impossibility of the desired assignment and hence proves the GKS theorem in dimension 2.

We now give a proof of the GKS theorem in $d = 2j+1$ dimensions, of which Nakamura's proof is a special case. Consider $n$ arbitrary directions in space and the $d(=2j+1)$ spin states of a spin-$j$ particle along each of these directions. For each spin state along each direction, introduce a POVM element that is the projector onto that state divided by the positive integer $r$. All the elements associated with any $r$ directions then constitute a POVM, and the total number of such POVMs is $N = {}^nC_r \equiv n!/\{r!(n-r)!\}$. The number of POVMs in which each element occurs is $M = {}^nC_r - {}^{n-1}C_r$. If $N$ is odd and $M$ is even, the GKS theorem can be established by the same *reductio ad absurdum* argument as in the $d=2$ case for, on the one hand, the total number of occurrences of elements with value 1 in all the POVMs is required to be odd (since there must be exactly one such element per POVM) while, on the other, it is also required to be even (since each such element is repeated twice over the POVMs). Nakamura's proof is the simplest case of this proof with $j = \frac{1}{2}, d = 2, n = 3, r = 2, N = 3$ and $M = 2$ and the three arbitrary directions chosen (unnecessarily) along the diameters of a regular hexagon.

Cabello's proof [1] can likewise be generalized by using ten arbitrary directions in space (instead of the ten threefold axes of a regular dodecahedron) and using all $d = 2j+1$ spin states of a spin-$j$ particle along each of these directions. Only the topology of a dodecahedron (and not any of its metrical properties) plays a role in this proof, since the POVMs continue to retain all their essential properties as the dodecahedron is deformed arbitrarily.

While the Cabello and Nakamura constructions in $d \geq 3$ prove the GKS theorem, they are not rich enough to yield proofs of the KS theorem, and there is no obvious way of altering them to achieve this goal. It is therefore interesting to look for other GKS proofs in $d \geq 3$ that can be extended into proofs of the KS theorem. We now present one such proof in $d = 4$.

Consider the three tesseracts (i.e. four dimensional hypercubes) that can be inscribed in a 24-cell. A 24-cell is a four-dimensional regular polytope with 24 vertices whose coordinates, relative to its center, can be taken as $(2,0,0,0), (0,2,0,0), (0,0,2,0), (0,0,0,2), (1,1,1,1), (1,-1,1,-1), (1,1,-1,-1)$, $(1,-1,-1,1), (1,1,1,-1), (-1,1,1,1), (1,-1,1,1)$ and $(1,1,-1,1)$, together with the negatives of these vectors. The first four vectors listed, together with their negatives, constitute the vertices of a cross polytope (or hyperoctahedron), as do the next four vectors, and the last four. By taking the vertices of these cross polytopes in pairs, we obtain the vertices of the three tesseracts that can be inscribed in the 24-cell. Note that two inscribed tesseracts meet at each vertex of the 24-cell. We now introduce 12 states (or rays) of a 4-state quantum system that are derived from the vertices of a 24-cell. We take the components of each ray, in the standard basis, to be proportional to the coordinates of one of the vertices of a 24-cell, with each pair of antipodal vertices counted only once. We further number the rays from 1 to 12 so that they correspond to the vertices of the 24-cell in the order given above. The rays divide into three groups (or "tetrads") of mutually orthogonal rays, $(1,2,3,4)$, $(5,6,7,8)$ and $(9,10,11,12)$, that each correspond to one of the cross polytopes inscribed in the 24-cell.

We next introduce as POVM elements the projectors onto the above 12 rays, each multiplied by ½, and note that: (a) the elements belonging to any two tetrads, and hence to one of the inscribed tesseracts, constitute a POVM, (b) there are 3 such POVMs in all, corresponding to the three tesseracts that can be inscribed in the 24-cell, and (c) each element belongs to exactly two POVMs, as a consequence of the fact that two inscribed tesseracts meet at each vertex of the 24-cell. The proof of the GKS theorem then follows from the same sort of parity argument as before, which can be rephrased here as the geometrical statement that it is impossible to color each antipodal pair of vertices of a 24-cell white or black in such a way that each inscribed tesseract has exactly one antipodal pair of vertices colored white. Note, however, that the present GKS proof, unlike the Cabello and Nakamura proofs, is a metrical (and not a topological) one, since any flexing of the 24-cell causes the POVMs (and hence the proof) to fall apart.

The 12 rays on which the above proof is based do not suffice for a KS proof, which requires that each ray be colored white or black in such a way that every orthogonal tetrad has exactly one white ray in it. However a KS proof can be generated by adding 12 further rays derived from the dual of the 24-cell considered, with the vertices of the dual having coordinates $(\pm 1, \pm 1, 0, 0)$ and all permutations thereof. The KS proof based on this set of 24 rays was given by Peres [5]. The author [6] pointed out that Peres' KS proof could be promoted into a proof of Bell's theorem [7] by using the correlations in a singlet state of two spin-$\frac{3}{2}$ particles to justify the assumption of noncontextuality made in the KS proof. Thus, this set of 24 rays suffices to prove the GKS, KS and Bell theorems. However this demonstration is not totally satisfying because some of the rays can be dispensed with in each of the individual proofs.

A more economical set of 18 rays can be obtained from the above set as follows [8]. First number from 13 to 24 the rays associated with the vertices $(1,0,1,0), (0,1,0,1), (1,0,-1,0), (0,1,0,-1)$, $(1,1,0,0), (1,-1,0,0,), (0,0,1,1), (0,0,1,-1), (1,0,0,1), (0,1,1,0), (1,0,0,-1)$ and $(0,1,-1,0)$ of the dual 24-cell. The 18 rays then result on omitting rays 1,5 and 10 from the first set of twelve and rays 16,20 and 24 from the second set of twelve. These 18 rays form the nine orthogonal tetrads shown in Table 1, with each ray occurring in exactly two tetrads. The oddness of the total number of tetrads, together with the evenness of the occurrence of each ray among the tetrads, then allows the KS theorem to be proved using the same sort of parity argument as before [8]. A proof of the GKS theorem can be obtained by introducing POVM elements that are the projectors onto these 18 rays, each multiplied by a factor of 1/3. Then it seen that the elements in tetrads T1,T5 and T7 form a POVM, as do the elements in T2,T4 and T8, and the elements in T3,T6 and T9. Further, each element occurs in exactly two POVMs. The two preceding facts suffice to establish the GKS theorem via the usual parity argument. It has been demonstrated [9] that the 18-ray KS proof can be converted into a Bell proof by making use of the entanglement afforded by a pair of Bell states. Thus this set of 18 rays, like the previous set of 24, also suffices for a proof of all three theorems.

But what makes this 18-ray set more interesting is that it appears to be a "critical" set for the proofs of all three theorems in the sense that deleting even a single ray (or POVM element) from it causes each of the three proofs to collapse. The criticality of the KS proof was demonstrated in the second paper of Ref.[8] and the criticality of the Bell proof, which rests directly on the KS proof, follows as a corollary. The criticality of the GKS proof appears highly plausible, although we do not have an ironclad proof of it yet.

Several other KS-Bell proofs, such as the ones based on the 40 rays of the "Penrose dodecahedron" or the 60 rays of the 600-cell [6], can be made to yield GKS proofs by suitably regrouping the projectors involved in the KS proof into POVMs [11]. However none of these proofs is as economical as the 18-ray proof given here, and none is "critical" in the sense defined above.

It has been demonstrated [6,10,12] that *any* proof of the KS theorem can be turned into a proof of Bell's theorem by making use of the right kind of entanglement. This strong KS-Bell connection leads one to ask whether a similar KS-GKS connection exists as well. The example given in this paper, along with a few others [10], would seem to suggest that the projectors occurring in any KS proof can be turned, with sufficient ingenuity, into POVM elements that furnish a GKS proof. A proof of this conjecture, or a counterexample to it, would be interesting.

**Acknowledgement**. I thank A.Cabello for his helpful comments on an earlier draft of this paper.

REFERENCES

[1] A.Cabello, Phys. Rev. Lett. **90**, 190401 (2003).
[2] S.Kochen and E.P.Specker, J. Math. Mech. **17**, 59 (1967); J.S.Bell, Rev.Mod.Phys. **38**, 447 (1966).
[3] A.M.Gleason, J. Math. Mech. **6,** 885 (1957). The generalization of Gleason's theorem to the case of POVMs is discussed in P.Busch quant-ph/9909073, C.A.Fuchs quant-ph/0205039 and C.M.Caves, C.A.Fuchs, K.Manne and J.Renes (unpublished ).
[4] The "elements" of a POVM are sometimes referred to as "effects", as in the book by K.Kraus, *States, effects, and operations*, Springer-Verlag, Berlin, 1983.
[5] A.Peres, J.Phys. **A24,** 174 (1991). Important precursors to this paper were A.Peres, Phys.Lett. **A151**, 107 (1990) and N.D.Mermin, Phys.Rev.Lett. **65**, 3373 (1990).
[6] P.K.Aravind, Phys. Lett. **A262**, 282 (1999). This paper shows how an arbitrary proof of the KS theorem in dimension $d = 2j+1$ can be converted into a proof of Bell's theorem by making use of the correlations in a singlet state of two spin-$j$ particles.
[7] J.S.Bell, Physics *1*, 195 (1964).
[8] A.Cabello, J.M.Estebaranz, and G.Garcia-Alcaine, Phys. Lett. **A212**, 183 (1996); P.K.Aravind, Found. Phys. Lett. **13**, 499 (2000).
[9] P.K.Aravind and V.C.Babau, quant-ph/0104133v1. Instructions are given for measuring the observables whose nondegenerate eigenstates are the tetrads T1-T9 of Table 1.
[10] P.K.Aravind, Found. Phys. Lett. **15**, 397 (2002). This paper shows how an arbitrary proof of the KS theorem in dimension $d = 2^n$ can be converted into a proof of Bell's theorem by making use of the correlations in $n$ Bell states shared symmetrically by two observers.
[11] P.K.Aravind (unpublished).
[12] The KS-Bell connection was established in specific cases in P.Heywood and M.L.G.Redhead, Found. Phys. **13**, 481 (1983), J.Zimba and R.Penrose, Stud. Hist. Phil. Sci. **24**, 697 (1993), N.D.Mermin, Rev. Mod. Phys. **65**, 803 (1993) and A.Cabello, Phys. Rev. Lett. **87**, 010403 (2001).

| T1 | 2  3  21  23 | T2 | 2  4  13  15 | T3 | 3  4  17  18 |
|----|--------------|----|--------------|----|--------------|
| T4 | 6  7  21  22 | T5 | 6  8  17  19 | T6 | 7  8  13  14 |
| T7 | 9  11  14  15 | T8 | 9  12  18  19 | T9 | 11  12  22  23 |

TABLE 1. The nine tetrads, T1 through T9, formed by the 18 rays described in the text. Note that each ray occurs in exactly two tetrads.